\documentclass[aps,pra,amsmath,superscriptaddress,reprint,floatfix,notitlepage,balancelastpage,twocolumn,showpacs,reprint]{revtex4}
\usepackage[colorlinks,urlcolor=red,citecolor=red]{hyperref}
\usepackage{graphicx,subfigure} 
\usepackage{epstopdf}
\usepackage{amsthm} 

\usepackage{tikz}

\let\baraccent=\= 
\renewcommand{\=}[1]{\stackrel{#1}{=}} 

\theoremstyle{definition}

\theoremstyle{remark}

\newcommand{\xit}{\tilde{\xi}}
\newcommand{\lambdah}{\hat{\lambda}}

\newcommand{\curC}{{\cal C}}
\newcommand{\curH}{{\cal H}}

\newcommand{\phdag}{{\phantom{\dagger}}}
\newcommand{\ps}{{\phantom{*}}}

\newcommand{\real}{{\rm Re}}

\newcommand{\muh}{\hat{\mu}}

\newcommand{\psih}{\hat{\psi}}

\newcommand{\Psit}{\tilde{\Psi}}

\newcommand{\br}{{\bf r}}

\newcommand{\be}{\begin{equation}}
\newcommand{\ee}{\end{equation}}
\newcommand{\bea}{\begin{eqnarray}}
\newcommand{\eea}{\end{eqnarray}}
\newcommand{\bse}{\begin{subequations}}
\newcommand{\ese}{\end{subequations}}

\begin{document}

\title{Pairing correlations in a trapped one-dimensional Fermi gas}
\author{Stephen Kudla}
\author{Dominique M. Gautreau}
\author{Daniel E. Sheehy}
\email{sheehy@lsu.edu}
\affiliation{Department of Physics and Astronomy, Louisiana State University, Baton Rouge, LA, 70803, USA}
\date{March 19, 2015}
\begin{abstract} We use a BCS-type variational wavefunction to study
attractively-interacting quasi one-dimensional (1D) fermionic  atomic
gases, motivated by cold-atom experiments that access the 1D regime
using an anisotropic harmonic trapping potential (with
trapping frequencies $\omega_x = \omega_y \gg \omega_z$) that confines
the gas to a cigar-shaped geometry.  To handle the presence of the
trap along the $z$-direction, we construct our variational wavefunction from the
harmonic oscillator Hermite functions that are the eigenstates of the
single-particle problem.
  Using an analytic determination of the
effective  interaction among harmonic oscillator states along with a
numerical solution of the resulting variational equations, we make
specific experimental predictions for how pairing correlations would be
revealed in experimental probes like the local density and the momentum
correlation function.  
\end{abstract}


\maketitle

\section{Introduction}
In recent years there has been much interest in the phenomena
 of pairing and superfluidity of trapped fermionic atomic gases\cite{BlochReview,Giorgini,Guan}.  
The questions being addressed by experiments with ultracold fermions
are quite general and concern the possible many-body phases of interacting fermions
as a function of experimentally controllable parameters such as temperature, 
interaction strength and the densities of various species of fermion. 

 For the case of
two species of fermions, relevant to the analogous problem of interacting spin-$\frac{1}{2}$ 
electrons in electronic materials, attractively interacting fermionic atomic gases 
are predicted to exhibit several interesting many-body phases.  These include a homogeneous
paired superfluid phase for equal densities of the two species (a \lq\lq balanced\rq\rq\ gas) 
that undergoes a crossover from
Bose-Einstein condensation (BEC) to Bardeen-Cooper-Schrieffer (BCS) pairing as a function 
of interfermion interactions, and a spatially-inhomogeneous Fulde-Ferrell-Larkin-Ovchinnikov~\cite{FF,LO}
(FFLO) superfluid for unequal densities of the two species (an \lq\lq imbalanced\rq\rq\ gas).  

Recent experiments~\cite{LiaoRittner} have explored two-species, attractively-interacting fermionic atomic gases 
in a quasi one-dimensional (1D) geometry, of interest since the 
regime of stability of the FFLO state is theoretically predicted~\cite{Orso07,HuLiuDrummond} to be much wider
than in the 3D case~\cite{Radzihovsky} (at least within the simplest mean-field approximation~\cite{RV}).  These experiments showed a remarkable quantitative agreement 
between experiment and theory for the local densities $n_\sigma(z)$ of the two species ($\sigma = \uparrow,\downarrow$)
of atoms within a theoretical approach that combined exact Bethe-Ansatz analysis of an infinite 1D gas with
the local density approximation (LDA) to handle the spatial variation of the trap.

If the imbalanced superfluid phase of 1D fermion gases 
posseses FFLO-type pairing correlations (as indicated theoretically~\cite{Feiguin,LiuPRA2007,Batrouni,HM2010,Sun,SunBolech}), and if the LDA holds (so that 
the uniform case phase diagram is relevant for a trapped gas), then trapped 1D imbalanced Fermi gases
may provide the best opportunity to observe signatures of the FFLO state.

\begin{figure}[ht!]\vspace{-.25cm}
\centering
\includegraphics[width=80mm]{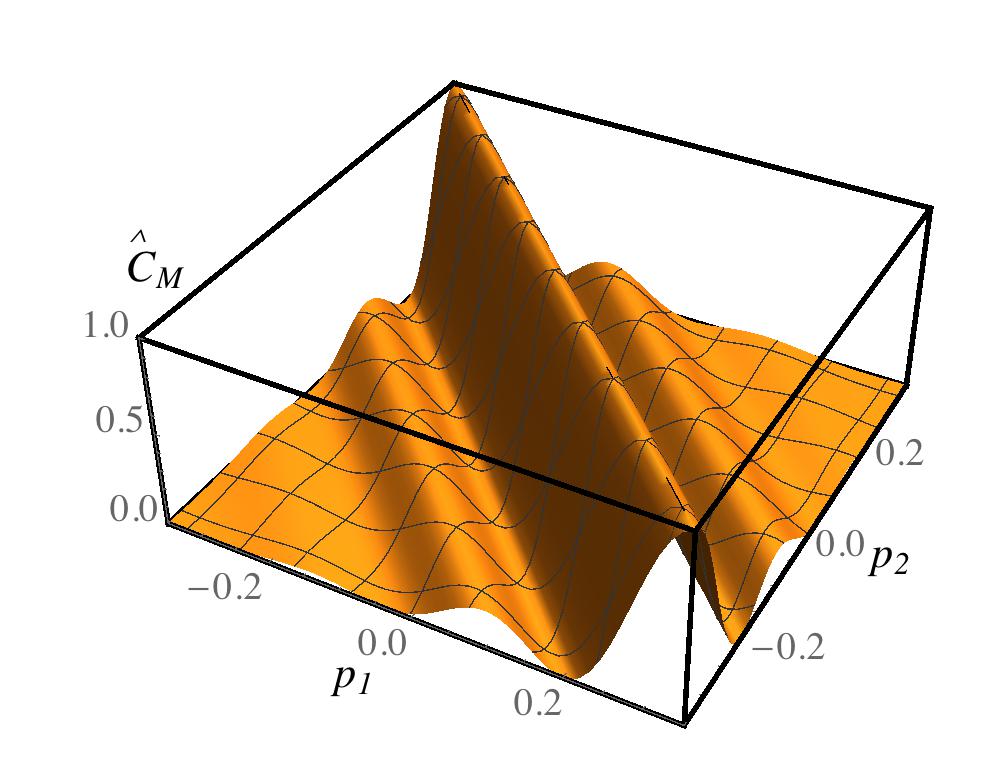}\vspace{-.25cm}
\caption{(Color Online) The in-trap momentum correlation function $\curC_M(p_1,p_2) = \langle n_{p_1\uparrow}n_{p_2\downarrow}\rangle - 
\langle n_{p_1\uparrow}\rangle \langle n_{p_2\downarrow}\rangle$ of a balanced 1D trapped fermion gas
with attractive interactions, normalized to its value at $p_1=p_2=0$ (which we call $\hat{\curC}_M(p_1,p_2)$).  This observable, 
which can be experimentally accessed
by a measurement of noise correlations after free expansion of the gas~\cite{Altman04}, shows a
rapid dependence on the sum of the momenta $p_1 + p_2$ (while being almost 
independent of $p_1-p_2$), providing a probe of fermionic pairing correlations.}
\label{fig:one}
\end{figure}

However, a central outstanding question concerns how to directly probe pairing correlations in this system.  
To address this, we have studied trapped attractively-interacting Fermi gases in 1D using a variational
wavefunction that accounts for the remaining trapping potential along the $z$-direction without
using the LDA.  
Our variational wavefunction is of the BCS form, but in which the pairing is among harmonic oscillator
states.  An additional variational parameter of our wavefunction, beyond the usual BCS coherence
factors, is the effective oscillator length of the single-particle states in our basis, which is allowed
to be distinct from the actual oscillator length associated with the trapping potential.  As discussed below,
including this parameter is necessary for the wavefunction to have a physically sensible 
density profile, and we find a density profile that agrees reasonably well with Bethe Ansatz combined with 
LDA.

The purpose of this paper is to study pairing and interaction effects in quasi-1D trapped fermionic
atomic gases.  We find that a striking probe of pairing correlations is the in-trap momentum
correlation function $\curC_M(p_1,p_2) = \langle n_{p_1\uparrow}n_{p_2\downarrow}\rangle - 
\langle n_{p_1\uparrow}\rangle \langle n_{p_2\downarrow}\rangle
$ with $p_1$ and $p_2$ momenta along the $z$-direction and
$n_{p\sigma}$ the occupation of the state with momentum $p$ and spin $\sigma$.
This quantity can be determined experimentally by measuring real space atom density
 correlations after free expansion of 
the gas, as shown theoretically~\cite{Altman04}
 and implemented experimentally~\cite{Greiner05} in the 3D case.  In the present 1D case, the free expansion
would be along the tube axis after lowering the confining potential along the $z$-direction.

Other recent theoretical work has analyzed expansion of 1D gases, focusing primarily on the imbalanced
case~\cite{Lu,Bolech12}.  Since our interest is understanding pairing correlations, here we study the balanced gas, using
a novel oscillator-basis approach.  A similar method can apply to imbalanced gases (relevant for the FFLO state), which we will
present in a future work~\cite{DGpaper}.
In Fig.~\ref{fig:one} we plot our results for the normalized quantity
$\hat{\curC}_{\rm M}(p_1,p_2) \equiv \curC_M(p_1,p_2)/\curC_M(0,0)$ showing a strong dependence on $p_1$ and $p_2$, that
we can connect to the nature of the underlying pairing correlations as discussed below. 

This paper is organized as follows.  In Sec.~\ref{sec:MH} we introduce the system Hamiltonian, describe our pairing wavefunction 
in the oscillator basis, and explain why we must allow the oscillator length associated with our single-particle states
to differ from the trap oscillator length.  In Sec.~\ref{sec:ve} we compute the variational ground-state energy and provide an analytic result
for the effective interaction function appearing in this energy.  In Sec.~\ref{sec:vareqs}, we present the equations
that come from minimizing the variational ground-state energy.  In Sec.~\ref{sec:res}, we describe our numerical solutions to these variational
equations which yield our predictions for  the local density and local pairing potential.  
  In Sec.~\ref{sec:momcorr}, we describe how we obtain the momentum correlation function for a trapped 1D fermion gas and obtain an approximate
analytic formula for this quantity.  In Sec.~\ref{sec:ba}, we analyze our system using the Bethe ansatz along with the local 
density approximation, with the comparison to our variational method given in Fig.~\ref{fixedden}.  Finally, we conclude in Sec.~\ref{sec:cr}.

\section{Model Hamiltonian and variational wavefunction}
\label{sec:MH}
Our starting point is a Hamiltonian for 
an attractively interacting fermion gas confined to a harmonic trap 
$V(\br) =  \frac{1}{2} m \big[\omega_\perp^2 \rho^2 + \omega_z^2z^2\big]$, with $\omega_\perp\gg\omega_z$,
such that, at sufficiently low fermion density, we can restrict attention to the lowest 
oscillator level associated with $\omega_\perp$.
 The resulting quasi one-dimensional Hamiltonian, with $\Psi_\sigma(z)$  the field operator for spin-$\sigma$, is: 
\bea
\nonumber 
&&\curH = 
\int_{-\infty}^\infty dz\,\Big( \sum_{\sigma = \uparrow,\downarrow}\Psi_\sigma^\dagger(z)\Big[\frac{p_z^2}{2m} 
+V(z) 
\Big]\Psi_\sigma(z) 
\\
&&\qquad \qquad 
+ \lambda \Psi_\uparrow^\dagger(z)  \Psi_\downarrow^\dagger(z)  \Psi_\downarrow^\phdag(z)  \Psi_\uparrow^\phdag(z) \Big),
\label{eq:modelham}
\eea
where  $V(z) = \frac{1}{2}m\omega_z^2 z^2$ is the trap along the $z$-direction,  and 
the coupling parameter  is $\lambda = -2\hbar^2/ma_{1D}$ with $a_{1D}$ the one-dimensional scattering
length~\cite{Olshanii98}.  We proceed by expressing $\Psi_\sigma(z)$  in terms of harmonic oscillator eigenfunctions $\psi_n(z)$ (with $n=0,1,\cdots$ the oscillator level index) via:
\bea
\Psi_\sigma(z) &=& \sum_n \psi_n(z) a_{n\sigma}, 
\\
\label{eq:hoe}
\psi_n(z) &=& \frac{1}{\sqrt{2^n n!a_z}} \frac{1}{\pi^{1/4}} {\rm e}^{-z^2/2a_z^2} H_n(z/a_z),
\eea
with  $H_n(z)$ the Hermite polynomial and 
\be
\label{Eq:ol}
a_z = \sqrt{\frac{\hbar}{m\omega_z}},
\ee
  the the oscillator length.
The operator  $a_{n\sigma}$ annihilates a fermion with spin $\sigma$ in the $n$th harmonic oscillator level
with single-particle energy $\epsilon_n = \hbar \omega_z(n+\frac{1}{2})$.
The system Hamiltonian in this basis is, defining $\hat{\curH}   = \curH/\hbar\omega_z$,
\be
\label{Eq:inbasis}
\hat{\curH} =  \sum_{n,\sigma} \hat{\epsilon}_n   a_{n_\sigma}^\dagger  a_{n_\sigma}^\phdag+\lambdah
   \sum_{n_i}\lambda_{\{n_i\}}  
a_{n_1\uparrow}^\dagger a_{n_2\downarrow}^\dagger  a_{n_3\downarrow}^\phdag a_{n_4\uparrow}^\phdag ,
\ee
where the normalized single-particle energy is $\hat{\epsilon}_n = \epsilon_n/\hbar\omega_z$.
Here, $\lambda_{\{n_i\}}$ is shorthand for 
\be
\lambda_{n_1,n_2,n_3,n_4}   \equiv  \int_{-\infty}^\infty dz \, 
 \psi_{n_1}(z) \psi_{n_2}(z) \psi_{n_3}(z) \psi_{n_4}(z),
\label{eq:integral}
\ee
characterizing interactions among the oscillator states, and we have 
 have introduced 
\be
\label{Eq:lambdahatdef}
\lambdah = \frac{\lambda}{\hbar\omega_z a_z} = -2\frac{a_z}{a_{1D}},
\ee
the  dimensionless coupling parameter.

\begin{figure}[ht!]
     \begin{center}
        \subfigure{
            \label{fig:first}
            \includegraphics[width=85mm]{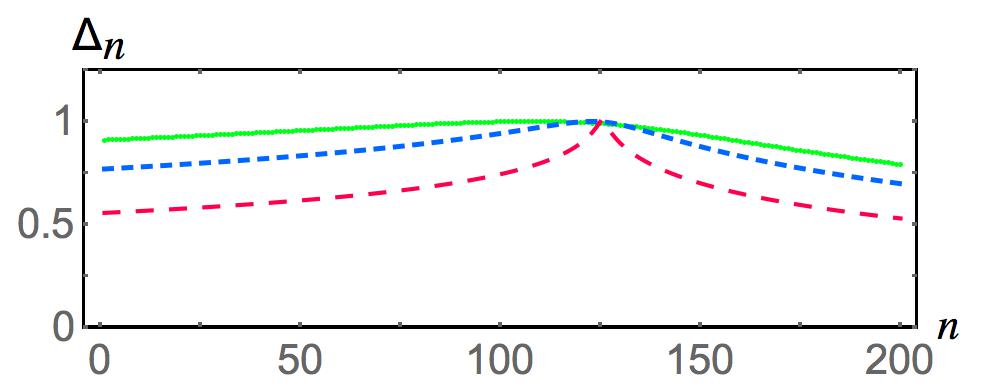}}
        \\ \vspace{-.25cm}
        \subfigure{
            \label{fig:fourth}
            \includegraphics[width=85mm]{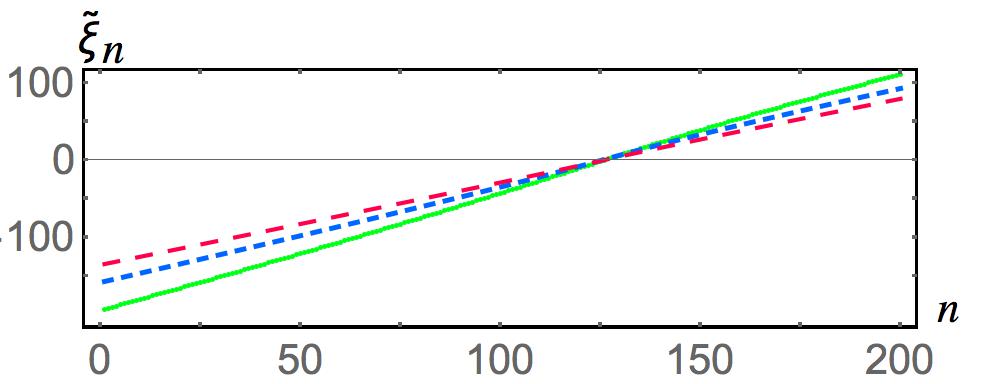}}
    \end{center}\vspace{-.5cm}
    \caption{
        (Color Online) The top panel shows the pairing amplitude for harmonic oscillator level $n$ (normalized to its
peak value) and the bottom panel shows the  renormalized dispersion $\xit_n$ for $\lambdah = -4.8$ (red, long dashing), 
$\lambdah = -13.6$ (blue, short dashing) and $\lambdah = -22.8$ (green solid) with the total particle number fixed at 
$N=250$.
     }\vspace{-.25cm}
   \label{fig:two}
\end{figure}

In the absence of interactions, the ground state of $\curH$ is simply
a Fermi gas in the oscillator basis, with harmonic oscillator  levels
$n\leq n_F$ occupied and $n>n_F$ empty.  A physically
sensible wavefunction that has this limiting case, but which also includes
the possibility of pairing correlations among single-particle states, is the following
BCS-type variational wavefunction:
\be
\label{eq:bcsform}
|\Psi\rangle = \prod_n (u_n + v_n a_{n\uparrow}^\dagger a_{n\downarrow}^\dagger)|0\rangle, 
\ee
where the coherence factors $u_n$ and $v_n$ satisfy the constraint $|u_n|^2+|v_n|^2=1$.  A trapped quasi 1D
Fermi gas is not expected to exhibit long-range pairing order. Thus, Eq.~(\ref{eq:bcsform}) should
break down on long length scales due to the absence of long-range phase coherence.  
 However, this wavefunction can capture {\em local\/} pairing correlations and their impact on 
observables like the local density and density-density correlations in a trapped gas.
An important task, that we leave for future work, is the investigation of how fluctuations around
our variational solution will modify our predictions.  For now, our goal is to understand the experimental
predictions of Eq.~(\ref{eq:bcsform}).

Before proceeding, however, we note that a crucial drawback of our ansatz, Eq.~(\ref{eq:bcsform}), is that it
yields a density profile corresponding to an atom cloud that {\em increases\/} in size along the axial direction in response to
increasing attraction.  To see the reason for this physically incorrect behavior, consider the noninteracting
($\lambdah\to 0$)  exact ground state, which is a Fermi gas with oscillator states filled up to the Fermi level $n_{\rm F}$.
Since the spatial extent of the harmonic oscillator wavefunction at level $n$
 is $\simeq a_z \sqrt{n}$, we can estimate the cloud size to be
approximately proportional to $\sqrt{n_{\rm F}}$ (fixed by the largest filled level). 
 If we now turn on attractive interactions, the Pauli principle
means that levels with $n <n_{\rm F}$ cannot increase their occupation, and that 
oscillator levels with $n>n_{\rm F}$, which have a larger spatial extent, will have a finite amplitude to become
occupied.  
The occupation of such higher levels of course does not 
imply a spatially larger cloud, since the local axial  density operator, expressed in the oscillator basis,
\be
\label{Eq:nhatgeneral}
\hat{n}(z) = \sum_{n,m,\sigma} \psi^*_n(z)  \psi_m(z) a_{n\sigma}^\dagger a_{m\sigma}^\phdag, 
\ee
has terms that are off-diagonal in the oscillator level.  In the true ground state, these off-diagonal terms
can lead to cancellations among the terms in Eq.~(\ref{Eq:nhatgeneral}), describing a 1D atomic gas that 
shrinks with increasing attractive interactions. 

However, the approximate BCS wavefunction Eq.~(\ref{eq:bcsform})
projects out such off-diagonal terms, yielding the expectation value $n(z) = \langle \Psi|\hat{n}|\Psi\rangle$ given by:
\be
\label{densityformula2015}
n(z) = 2\sum_{m=0}^\infty\, |\psi_n(z)|^2 |v_n|^2,
\ee
which will clearly exhibit a increased cloud size with increasing attractive interactions as higher oscillator levels become occupied, 
since all terms in the sum are positive.  

Remedying this physically incorrect behavior of our variational wavefunction is crucial, since the axial density is a primary
observable in cold atom experiments.   However, we aim to do this in a way that preserves the simplicity of
our BCS variational wavefunction.
 To accomplish this, we introduce an additional variational parameter, which is
the oscillator length associated with our wavefunctions, by replacing $a_z\to a$  in Eq.~(\ref{eq:hoe}) and considering
$a$ to be a variational parameter to be minimized.
Thus, while the noninteracting fermion gas occupies
oscillator states with an oscillator length that is related to the trap potential via Eq.~(\ref{Eq:ol}), in the interacting case the 
optimal (lowest energy) BCS-type state may involve oscillator states with $a<a_z$ that is smaller, allowing the 
cloud to shrink in spatial extent.  We therefore introduce the parameter 
\be
\label{Eq:def}
\eta = \frac{a_z^2}{a^2},
\ee
where $a_z$ remains the true oscillator length.  Note that we can also write $\eta = \omega/\omega_z$ with 
$\omega$ the frequency of a ficticious trap for which $ a$ is the oscillator length.  Then,
it is convenient to
split the trap potential into two pieces, via $V(z) = \frac{1}{2}m\omega^2 z^2 +  \frac{1}{2}m(\omega_z^2-\omega^2) z^2$,
where the first term yields a contribution to $\curH$ that is identical to Eq.~(\ref{Eq:inbasis}) but with 
$\omega_z\to \omega$ and the second term yields a correction that we will evaluate using the properties of the 
oscillator wavefunctions. 
  We find, upon repeating the preceding analysis 
   for the case of $\eta \neq 1$, the effective Hamiltonian 
\bea
\nonumber
&&\hat{\curH} = \eta \sum_{n,\sigma} \hat{\epsilon}_n   a_{n_\sigma}^\dagger  a_{n_\sigma}^\phdag+\sqrt{\eta}\lambdah
   \sum_{n_i}\lambda_{\{n_i\}}  
a_{n_1\uparrow}^\dagger a_{n_2\downarrow}^\dagger  a_{n_3\downarrow}^\phdag a_{n_4\uparrow}^\phdag 
\\
&& \hspace{-.5cm} +\frac{1}{2} \Big( \frac{1}{\eta} - \eta\big) 
\int_{-\infty}^\infty dz  \, z^2 \sum_{n_1,n_2,\sigma} \psih_{n_1}^*(z) \psih_{n_2}(z) a_{n_1\sigma}^\dagger a_{n_2\sigma}^\phdag,
\label{Eq:repeated}
\eea
with the second line coming from the abovementioned correction.  Here, 
 $\psih_n(z)$ is a dimensionless Hermite function (Eq.~(\ref{eq:hoe}) but with $a_z\to 1$) and we have once
again normalized to $\hbar \omega_z$ (as in Eq.~(\ref{Eq:inbasis})).

To summarize this section, Eq.~(\ref{Eq:repeated}) is an expression of our system Hamiltonian in terms
of creation and annihilation operators, $a_{n\sigma}$ and $a_{n_\sigma}^\dagger$, that correspond to harmonic
oscillator states with oscillator length $a$ that is different from the physical oscillator length of 
our system. 
Here and below $a$ generally appears only via the parameter $\eta$ Eq.~(\ref{Eq:def}), and we typically normalize all length scales to
$a_z$ and all energy scales by $\hbar \omega_z$ (for example in figures).
 Next we proceed by assuming that these oscillator states undergo pairing correlations
described by Eq.~(\ref{eq:bcsform}) and determine the optimal coherence factors and value of $a$.

\section{Variational Energy}
\label{sec:ve}
Upon taking the expectation value of the Hamiltonian using the wavefunction Eq.~(\ref{eq:bcsform}), in the second
line of Eq.~(\ref{Eq:repeated}) the only nonzero contribution comes from $n_1 = n_2$, allowing the $z$ integral to be
easily evaluated.  We then find that the normalized grand free energy $E_G= \langle \hat{\curH} - \muh \hat{N} \rangle$ 
(with $\hat{N}$ the number operator,
and $\muh= \mu/\hbar \omega_z$ the normalized chemical potential) is:
\be
\label{Eq:eeegeeeoned}
E_G = 2\sum_n \xi_n |v_n|^2 +\sqrt{\eta}\lambdah \sum_{n,m}\lambda_{n,m} \big(  u_{n}^* v_{n}^\ps
  v_{m}^* u_{m}^\ps+ |v_{n}|^2|v_{m}|^2\big),
\ee
where we defined $\xi_n = \frac{1}{2}\big(\eta + \eta^{-1}\big)n+\frac{1}{2} -\muh$.  Here and below we focus on zero temperature.
In the interaction part of Eq.~(\ref{Eq:eeegeeeoned}),
 the first term corresponds to pairing correlations and the second term corresponds to Hartree-Fock 
correlations.  Here, $\lambda_{m,n} \equiv  \lambda_{n,n,m,m}$ is the effective interaction resulting from our
variational ansatz, explicitly given by:
\be
\lambda_{m,n} = \frac{1}{2^{m+n}}\frac{1}{\pi n!m!} \int_{-\infty}^{\infty}
 dz \,{\rm e}^{-2z^2}
H_n^2(z) H_m^2(z).
\label{lambdamnresult}
\ee
Integrals of this form have been of interest to the mathematical
physics community~\cite{Wang}, and have also recently appeared in other cold-atom contexts~\cite{Rey}.  
Although it can be evaluated numerically, this becomes 
difficult for large $m$ and $n$.  We next present our analytic result for Eq.~(\ref{lambdamnresult}),
that greatly sped-up our calculations.
  To do this we
use an identity for the square of a Hermite polynomial, 
$H_n^2(z) = 2^n (n!)^2 \sum_{s=0}^n \frac{H_{2s}(z)}{2^s (s!)^2 (n-s)!}$,
for the two factors $H_n^2(z)$ and $ H_m^2(z)$ in Eq.~(\ref{lambdamnresult}).  This leads to  
 a $z$-integral involving a product of two Hermite polynomials
multiplying the Gaussian factor ${\rm e}^{-2z^2}$ that appears in Gradshteyn and Ryzhik~\cite{Gradshteynintegral,Gradshteyn}.
Then, evaluating the remaining summations, we obtain:
\be
 \lambda_{m,n}=\frac{(-1)^m}{\sqrt{2}m!}
\frac{_3F_2(\frac{1}{2},\frac{1}{2},-n; 1, \frac{1}{2}-m; 1\big)}{\Gamma[\frac{1}{2}-m]},
\label{Eq:couplingresult}
\ee
with $_3F_2$ the generalized hypergeometric function.  Although it is not obvious from
Eq.~(\ref{Eq:couplingresult}), $\lambda_{m,n}$ is indeed symmetric
under interchange of its indices.
%

\section{Variational equations}
\label{sec:vareqs}
We now proceed with minimizing  Eq.~(\ref{Eq:eeegeeeoned}) with respect to our variational parameters. 
To minimize with respect to the $u_n$ and $v_n$, we must
enforce the constraint  $|u_n|^2+|v_n|^2=1$ with a  Lagrange multiplier $E_n$. The resulting Euler-Lagrange equations take 
the form of a Bogoliubov-de Gennes (BdG) eigenvalue problem
\be
\begin{pmatrix} \xi_n + U_n & \Delta_n \\
\Delta_n^* & -\xi_n - U_n\end{pmatrix} \begin{pmatrix}u_n \\ v_n \end{pmatrix} = E_n  \begin{pmatrix}u_n \\ v_n \end{pmatrix},
\ee
where we defined 
 the strength of pairing correlations $\Delta_n \equiv - \lambdah\sqrt{\eta}\sum_{m} \lambda_{n,m} v_m^* u_m$ and the
 Hartree-Fock energy shift $U_n \equiv \lambdah\sqrt{\eta}\sum_m \lambda_{n,m} |v_m|^2$.  Defining 
the renormalized single-particle energy $\xit_n = \xi_n +U_n$, we obtain the BdG solution 
$u_n = \frac{1}{\sqrt{2}}\sqrt{1+\frac{\xit_n}{E_n}}$ and $v_n = \frac{1}{\sqrt{2}}\sqrt{1-\frac{\xit_n}{E_n}}$,
with $E_n = \sqrt{\xit_n^2 +\Delta_n^2}$.  Inserting these solutions into the definitions of $\Delta_n$ and
$U_n$ then leads to the self-consistency conditions
\bse
\label{eq:udelta2}
\bea
\label{Eq:gapfinharm}
\Delta_n &=& -\lambdah\sqrt{\eta}\sum_{m=0}^\infty \lambda_{n,m} \frac{\Delta_m}{2E_m},
\\
U_n &=& \lambdah\sqrt{\eta}\sum_{m=0}^\infty \lambda_{n,m}\frac{1}{2}\big(1-\frac{\xit_m}{E_m} \big) .
\label{Eq:ufinharm}
\eea
\ese
A third variational equation comes from minimizing $E_G$ with respect to the parameter $\eta$ that determines
the optimal oscillator length characterizing our basis set.  We find, differentiating $E_G$ with respect to 
$\eta$, 
\bea
\label{eq:thirdequation}
  &&0 = \sum_n |v_n|^2 \big(n+\frac{1}{2}\big)\big(1 - \frac{1}{\eta^2}\big)
\\
&&\qquad + \frac{1}{2\sqrt{\eta}} \lambdah \sum_n \lambda_{n,m} \big(u_n^*v_nv_m u_m + |v_n|^2|v_m|^2\big).\nonumber
\eea
We see that the parameter $\eta$ multiplies $\lambdah$ in  Eqs.~(\ref{eq:udelta2}) determining the $u_n$ and $v_n$.  Since
we expect $\eta>1$ in equilibrium, this implies an effectively larger coupling in equilibrium, consistent with the
picture of the central density increasing due to the presence of attractive interactions. 

%
\begin{figure}[ht!]
\centering
\includegraphics[width=85mm]{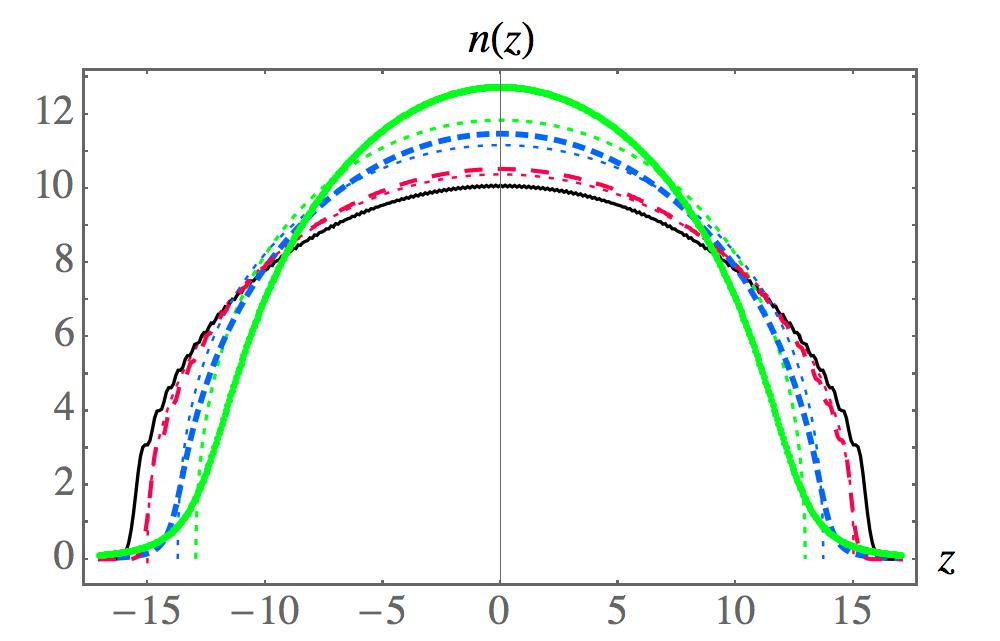}\vspace{-.25cm}
\caption{(Color Online)
The thick lines show the axial density $n(z)$ (in dimensionless units) resulting from our variational approach, as a function of position (normalized to the inverse
 oscillator length) for the same
parameters as Fig.~\ref{fig:two} (total particle number $N=250$ with $\lambdah = -4.8$ being red, long dashing; $\lambdah = -13.6$ being blue, short dashing,
and  $\lambdah = -22.8$ solid green).  Each such curve has a corresponding nearby dotted curve (with the same color scheme 
that, for $z=0$, is just below the variational result)
that is the result of Bethe ansatz along with the local density approximation 
for the same values of the dimensionless coupling constant and particle number.   The solid thin black curve that is the lowest at $z=0$ is
the noninteracting case.
}
\label{fixedden}
\end{figure}

\section{Results}
\label{sec:res}
The simultaneous numerical solution of  Eqs.~(\ref{eq:udelta2}) and Eq.~(\ref{eq:thirdequation}), yielding the variational parameters
describing our system ($\Delta_n$, $U_n$, and $\eta$), was done numerically,  although an approximate analytic solution can be found
in the extreme weak-coupling limit $\lambdah\to 0$ as described below.  Our numerical calculations were conducted for three values of
the dimensionless coupling ($\lambdah = -4.8$, $\lambdah = -13.6$
and $\lambdah = -22.8$) with the particle number held at $N=250$ (requiring an adjustment of the system chemical potential).
For comparison, the coupling in Ref.~\cite{LiaoRittner}
was $\lambdah \simeq -52$.

For our numerical procedure we truncated the sums in Eqs.~(\ref{eq:udelta2}) at 
an upper cutoff $n_{\rm max}=350$ (outside the plotted range of Fig.~\ref{fig:two}).  An
estimate of the error involved in this truncation comes from the value of 
$|v_{n_{\rm max}}|^2 = (2.9\times 10^{-7},1.8\times 10^{-4},2.6\times 10^{-3})$ for 
$\lambdah = -4.8$, $\lambdah = -13.6$
and $\lambdah = -22.8$, respectively, which we argue to be negligible except perhaps
in the $\lambdah = -22.8$ case.  For this coupling, we fit $|v_n|^2$ to a power
law for $n$ close to $n_{\rm max}$, and obtained a better error estimate by extrapolating
this beyond $n_{\rm max}$ and determining the expected number of fermions in levels
above $n_{\rm max}$, which we find to be 
$\Delta N  \simeq 2\int_{n_{\rm max}}^\infty dn\, |v_n|^2 \simeq 0.5$, much smaller than the total
particle number.

In Fig.~\ref{fig:two} (top panel), we plot our numerical results for the pairing
amplitude ($\Delta_n$), normalized to its maximum value $\Delta_{\rm
max}$.  The maximum pairing amplitudes were $\Delta_{\rm
max} = 0.71$, $\Delta_{\rm
max} = 15.0$ and $\Delta_{\rm
max} = 57.1$, for the coupling values
$\lambdah = -4.8$, $\lambdah = -13.6$
and $\lambdah = -22.8$
 respectively,  with the corresponding 
equilibrium $\eta$ values being $\eta = 1.09$, 
$\eta=1.31$ and $\eta  = 1.67$, with the latter describing a cold-atom cloud that shrinks in the axial direction,
effectively occupying oscillator states with $a<a_z$.

The weakest coupling
$\lambdah = -4.8$ plot (red dashed) shows that $\Delta_n$ is narrowly peaked
near the Fermi level $n_F \approx 125$ (defined by when $\xit_n$ comes closest to zero), 
consistent with the general expectation that
pairing is strongest near $n_F$.  We can approximately derive this behavior analytically
in the weak coupling (small $|\lambdah|$) limit by noting that, in this limit, the sum on the right side of
Eq.~(\ref{Eq:gapfinharm}) is dominated by terms near $n_F$.  If we approximate $\eta\approx 1$, take the dispersion to have the
form $\xit_n  = n- n_F$, and keep only the term $n=n_F$ in the sum, we
obtain $\Delta_n = \frac{1}{2}|\lambdah| \lambda_{n,n_F}$, so that the
shape of $\Delta_n$ approximately reflects the shape of the coupling
function Eq.~(\ref{Eq:couplingresult}). 
While this result qualitatively
captures the $n$ dependence of the pairing amplitude, it is only quantitatively
valid for $|\lambdah| \ll 1$ and does not approximately describe our
results for any of the displayed coupling values.   With increasing
attraction, $\Delta_n$ broadens considerably as more levels 
participate in pairing, as seen by the $\lambdah = -13.6$ (blue, short-dashing) and
$\lambdah = -22.8$ (green, solid) curves of Fig.~\ref{fig:two} (top panel).

The bottom panel of Fig.~\ref{fig:two} shows the renormalized dispersion $\xit_n$ 
for the same three coupling values, showing that this quantity is 
approximately linear near $n_F$ for all coupling values but with a renormalized
slope, with $\xit_n  = \alpha (n- n_F)$ where $\alpha$ increases with increasing coupling
strength.

We now turn to the question of how interaction effects would be revealed in
experiments.  A natural observable accessible in cold atom experiments
is the axial density $n(z)$ as a function of position, given by Eq.~(\ref{densityformula2015})
above
and plotted in Fig.~\ref{fixedden} for the same three coupling values (thick, with the same
color and dashing scheme as in Fig.~\ref{fig:two}).  The thin dotted curves that are adjacent to
our variational wavefunction results (with the same color scheme) are  the results
of combining the Bethe ansatz with the local density approximation (for the same parameters and 
particle number, with details provided in Sec.~\ref{sec:ba}), and the lowest solid curve
is the noninteracting case.

This figure shows that our variational method agrees quantitatively with the Bethe ansatz plus LDA for
the weakest coupling $\lambdah = -4.8$ case.  Since both curves are clearly distinct from the noninteracting
case, this is not merely due to the fact that they are all in the noninteracting limit, although both the variational
method and the Bethe ansatz plus LDA methods agree with the noninteracting curve for smaller coupling (for example, $\lambdah \simeq -0.1$).

  Increasing
the magntitude of the coupling strength causes the cloud to shrink in size (as expected), although the discrepancy
between  our variational results and the Bethe ansatz plus LDA also increases.  
  We note that, a priori, it is not clear which theoretical method is more accurate since
both are approximate, although we expect the Bethe ansatz plus LDA to be more  accurate
in the limit of a more uniform local density (which, here, occurs for smaller $|\lambdah|$) 

\begin{figure}[ht!]
\centering
\includegraphics[width=85mm]{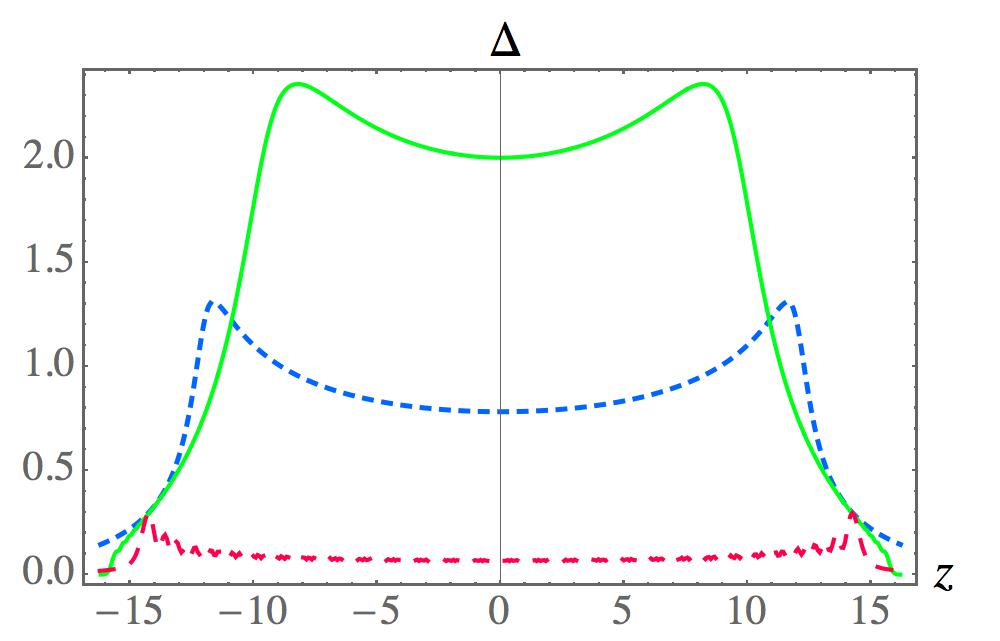}\vspace{-.25cm}
\caption{(Color Online) The local pairing amplitude $\Delta(z)$ (normalized to $\hbar \omega_z$), 
as a function of position (normalized to the oscillator length) for the same
parameters as Fig.~\ref{fig:two}, showing a significant increase in the local pairing with increasing attraction.}
\label{pairing}
\end{figure}

We now turn to the local pairing amplitude $\Delta(z) \equiv \langle \Psi_\uparrow(z) \Psi_\downarrow(z)\rangle$, given, 
within the present variational approach, by:
\be
\Delta(z)  = \sum_{n=0}^\infty \big(\psi_n(z)\big)^2 v_n u_n,
\ee
which we plot in Fig.~\ref{pairing}.  Strictly speaking, $\Delta(z)$ is not directly observable since it 
is off-diagonal in fermion field operators. However, it does provide information about the increasing strength
of pairing correlations with increasing magnitude of $\lambdah$.

 One way to estimate
the validity of our approach is to calculate the local BCS coherence length, given by $\xi = \frac{\hbar v_{\rm F}}{\pi \Delta}$
for a uniform system.  If $\xi$ is much larger than the typical interparticle spacing, then one expects  fluctuations around our
solution to be relatively small.  To determine this, we use the uniform-case result for the Fermi wavevector, $k_{\rm F} =\pi n/2$ (with $n$ the 1D atom density), in 
terms of which $v_{\rm F} =\hbar k_{\rm F}/m$.  Combining these gives 
\be
\label{xin}
\xi n = \hbar^2 n^2/(2m\Delta),
\ee
 for the coherence length normalized to the interparticle spacing $n^{-1}$.  The quantities on the right side, $n$ and $\Delta$, are plotted in 
Figs.~\ref{fixedden} and \ref{pairing}, but in dimensionless forms (normalized to $a_z^{-1}$ and $\hbar \omega_z$, respectively).  Converting 
the right side of this formula to dimensionless form yields $\hbar^2 n^2/(2m\Delta) \to n^2/(2\Delta)$ so that the normalized coherence length is 
simply the square of a curve in Fig.~\ref{fixedden} divided by a curve Fig.~\ref{pairing}.  Thus, we find $\xi n\agt 10$ for all coupling values,
indicating that the coherence length is large compared to the interparticle spacing.  This, along with
the approximate agreement with Bethe ansatz along with the LDA, gives further confidence in the validity of our approach.

 In the next section, we consider an observable, the momentum correlation function, which also probes the strength
of pairing correlations in a balanced 1D fermion gas.

\section{Momentum correlation function}
\label{sec:momcorr}
To find
a sensitive probe of pairing we turn to  the momentum correlation function 
$\curC_M(p_1,p_2) =\langle n_{p_1\uparrow}n_{p_2\downarrow} \rangle -\langle n_{p_1\uparrow}\rangle\langle n_{p_2\downarrow}\rangle$, 
with $n_{p\sigma} = c_{p\sigma}^\dagger c_{p\sigma}$ 
the momentum occupation operator.
As shown by Altman et al, $\curC_M(p_1,p_2)$ is probed by the real-space noise correlation
function $\curC(z_1,z_2) = \langle n_\uparrow(z_1) n_\downarrow(z_2)\rangle - \langle n_\uparrow(z_1)\rangle\langle n_\downarrow(z_2)\rangle$,
of the freely-expanded gas.  Thus, assuming the absence of interaction effects during expansion for time $t$, 
$\curC(z_1,z_2)$ is directly proportional to $\curC_M(p_1,p_2) $, with $p_1= mz_1/ t$ and $p_2= mz_2/t$.
Using our variational wavefunction, we find
$\curC_M(p_1,p_2)\propto |S(p_1,p_2)|^2$ with the sum
\be S(p_1,p_2)= \sum_{n=0}^\infty
\chi^\ps_{n}\big(p_1\big) \chi^\ps_n\big(p_2\big)  u_{n}^* v_{n}^\ps,
\label{correlatorsum}
\ee
where 
\bea
\chi_n(p) &=&  \int_{-\infty}^\infty dz\, {\rm e}^{-ipz} \psi_n(z), 
\\
&=& 
(-i)^n \frac{\pi^{1/4}\sqrt{a}}{\sqrt{2^{n-1} n!}} {\rm e}^{-a^2p^2/2} H_n(pa),
\eea
 are the Fourier-transforms of the harmonic oscillator wavefunctions (that we emphasize contain
$a$, the oscillator-length variational parameter).

\begin{figure}[ht!]
\centering
\includegraphics[width=80mm]{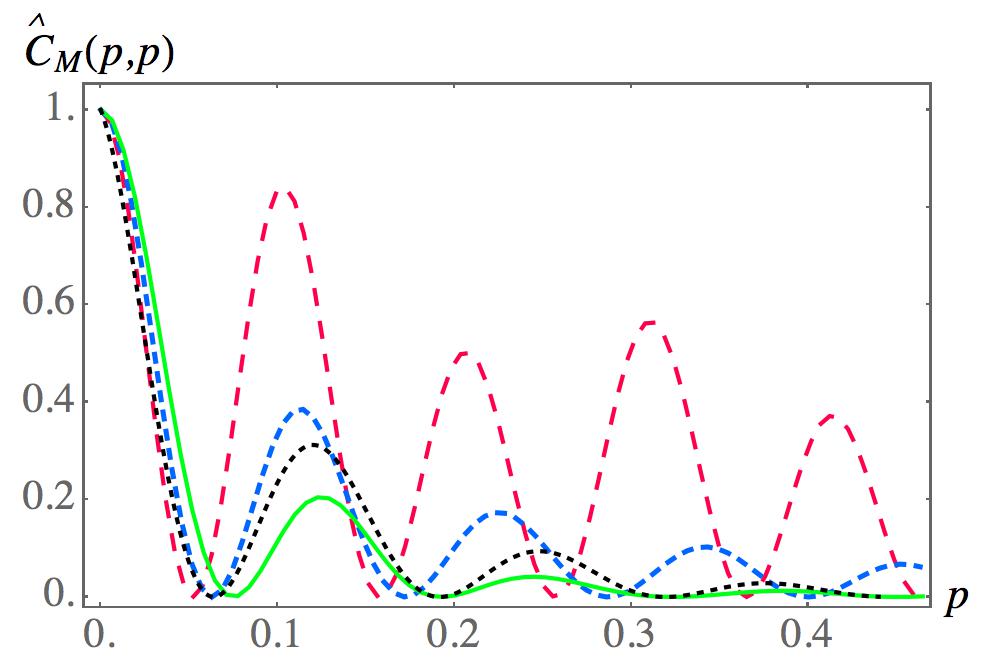}
\caption{(Color Online)
The normalized momentum correlation function, $\hat{\curC}_M(p_1,p_2)$  in the limit $p_1=p_2=p$ (normalized to unity at $p\to 0$), 
for a trapped 1D fermionic superfluid, for the same three coupling 
values as Figs.~\ref{fig:two} and \ref{fixedden}.   A fourth curve, black dots, depicts the approximate
theoretical formula Eq.~(\ref{Eq:chat}) for the case of $\lambdah = -22.8$ which should be compared to the 
green solid curve, our numerical result for this case.  Here we chose units for the momentum axis
such that $a_z = 1$.
}
\label{correlator}
\end{figure}

Our results for the momentum correlation function look, qualitatively, like Fig.~\ref{fig:one} for all
coupling values, where we plotted the normalized function $\hat{\curC}_M(p_1,p_2)\equiv \curC_M(p_1,p_2)/\curC_M(0,0)$.
Thus, $\hat{\curC}_M(p_1,p_2)$ is sharply peaked around $p_1+p_2=0$, and is, approximately, only a function of the sum 
$|p_1+p_2|$ of the momenta.  To understand the rapid variation as a function of $|p_1+p_2|$,
in Fig.~\ref{correlator}, we plot the equal momenta correlator $\hat{\curC}_M(p,p)$ for all three
coupling values (using the same color and dashing scheme as above), along with a fourth curve (black dots)
that is an approximate analytic evaluation of $\hat{\curC}_M(p,p)$ for the case of $\lambdah = -22.8$
that we now describe.

Our approximate form for the correlator follows by noting that the summand of Eq.~(\ref{correlatorsum}),
$u_{n}^* v_{n}^\ps  = \frac{\Delta_n}{2E_n}$,
is narrowly peaked for $n$ close to the Fermi level, with an
approximate Lorentzian shape for $n\to n_F$ given by $u_{n}^* v_{n}^\ps  \simeq
\frac{1}{2}\frac{1}{1+(n-n_F)^2/w^2},$
where  $w =
\sqrt{2}\Delta_{n_F}/\alpha$ approximately represents the number of
harmonic oscillator levels that are paired.  Here, we recall that $\alpha$ is the 
slope of the effective dispersion near the Fermi level, with $\xit_n = \alpha(n-n_F)$. From
the $\lambdah = -22.8$ results we find $\alpha \simeq 1.66$ and $\Delta_{n_F} \simeq  56.4$, 
yielding $w\simeq 48.0$.

 Expressing the Lorentzian 
in an integral form, $[1+(n-n_F)^2/w^2]^{-1}= w\int_0^\lambda d\lambda\, {\rm e}^{-\lambda w}\cos\lambda(n-n_F)$,
we can evaluate the sum to get (with $\real$ being the real part):
\be
S(p_1,p_2) = w\int_0^\infty d\lambda\, {\rm e}^{-\lambda w} \real\,\big( {\rm e}^{i\lambda n_F}
K\big[{\rm e}^{-i\lambda}\big]\big),
\ee
where $K[x]\equiv  \frac{\sqrt{\pi}}{\sqrt{1-x^2}}\exp\big[ \frac{(p_1^2+p_2^2)(1+x^2)+4p_1p_2x}{2(x^2-1)}\big]$.
The dominant contribution to this integral comes the regime where $\lambda\to 0$.  Expanding $K\big[{\rm e}^{-i\lambda}\big]$
in this limit yields an integral that can be easily evaluated analytically.  Finally taking the limit $w\ll n_F$ for
simplicity, we find for the normalized correlator: 
\be
\label{Eq:chat}
\!\!\!\hat{\curC}_{\rm M}(p_1,p_2)  = \cos^2 \big[ \sqrt{2n_F}|p_1+p_2|a\big] {\rm e}^{-\sqrt{2}w|p_1+p_2|a/\sqrt{n_F}},
\ee
which we find to be qualitatively accurate, as seen in Fig.~\ref{correlator}, connecting the local pairing and Hartree-Fock
correlations to this observable.  

As shown in Fig.~\ref{fig:one}, our numerical evaluation of $\hat{\curC}_{\rm M}(p_1,p_2)$ yields a result that is 
nearly independent of the difference in momenta $p_1-p_2$, a feature that appears in the approximate result 
Eq.~(\ref{Eq:chat}). However, we find that the degree to which $\hat{\curC}_{\rm M}(p_1,p_2)$ is independent of 
$p_1-p_2$ is rather sensitive to the choice of the upper cutoff $n_{\rm max}$ in our numerical summation,
and leave further investigation of this to future work. 

We also see that, within the approximations leading to Eq.~(\ref{Eq:chat}), the oscillatory variation of 
this correlation function as a function of the sum of momenta 
 measures the uppermost occupied
oscillator state (Fermi level $n_F$), and the exponential decay measures the strength of pairing 
at the Fermi level (via the parameter $w$). Thus, the momentum correlation function indeed provides
a direct probe of pairing correlations in a trapped 1D interacting Fermi gas. 

\section{Bethe ansatz and LDA}
\label{sec:ba}
In the present paper, our goal was to pursue a variational wavefunction scheme, based on a BCS type wavefunction 
in the oscillator basis, to analyze attractively interacting fermions in a one-dimensional trapping potential.  
In this section, we re-analyze our model Hamiltonian Eq.~(\ref{eq:modelham})
within a different approximation scheme, namely
the Bethe ansatz (exact for an infinite system, or $V(z)=0$) along with the local density approximation to handle the
trap.  Such a method was used in Ref.~\cite{LiaoRittner} in the imbalanced case and found to exhibit remarkable agreement 
with experimental results for the density profile.

To implement the Bethe ansatz, we follow the recent review of Guan et al~\cite{Guan}, taking the limit $\rho_1(k)=0$ of
Eqs.(13) of Ref.~\cite{Guan} (appropriate for the balanced case studied here).  Then, the density of pairs at 
quasimomentum $k$, $\rho(k)$, satisfies the Fredholm equation
\be
\label{baintegralequation}
\rho(k) = \frac{1}{\pi} +\int_{-A}^{A} dk'\, K(k-k') \rho(k') ,
\ee
with $K(x) = \frac{1}{\pi} \frac{c}{c^2+x^2}$, where $c$ is proportional to the 1D coupling constant (as defined below) 
The parameter $A$ is chosen so that the system has the correct total number of particles. 

To implement the Bethe ansatz, it is convenient to rescale coordinates in the Hamiltonian Eq.~(\ref{eq:modelham}) via 
$z \to a_zz$ with $a_z$ the oscillator length and define new fields $\Psi_\sigma(a_zz) = \Psit_\sigma(z)/\sqrt{a_z}$.  This leads to:
\bea
\nonumber 
&&\curH = 
\int_{-\infty}^\infty dz\,\Big( \sum_{\sigma }\Psit_\sigma^\dagger(z)\Big[\frac{p_z^2}{2ma_z^2} 
+\frac{1}{2} ma_z^2\omega_z^2 z^2
\Big]\Psit_\sigma(z) 
\\
&&\qquad \qquad 
+ \frac{\lambda}{a_z} \Psit_\uparrow^\dagger(z)  \Psit_\downarrow^\dagger(z)  \Psit_\downarrow^\phdag(z)  \Psit_\uparrow^\phdag(z) \Big),
\label{eq:hy}
\eea
which we see describes fermions of effective mass $m_{\rm eff} = ma_z^2$ and effective coupling $\lambda/a_z$.  Thus, while Guan et al
quote the relation $\lambda = \hbar^2 c/m$ between the coupling constant and the parameter $c$, in the present context
we should use this formula with the replacement $\lambda \to \lambda/a_z$ and $m \to ma_z^2$.  This leads to:
\be
c = - \frac{2a}{a_{1D}},
\ee
conveniently equal to our dimensionless parameter $\lambdah$ defined in Eq.~(\ref{Eq:lambdahatdef}).

Once we determine $\rho(k)$, via a numerical solution of Eq.~(\ref{baintegralequation}) for a chosen value of $A$, the 
 total particle number density $n$ and the dimensionless internal energy density are given by~\cite{Guan}:
\bea
n &=& 2 \int_{-A}^A dk\, \rho(k) ,
\\
\hat{E} &=&  \int_{-A}^A dk\,(2k^2-c^2/2) \rho(k) .
\eea
Note that, to obtain the system chemical potential, we need the dimensionful energy  density $E = \hbar^2/(2m_{\rm eff})\hat{E} $,
in terms of which $\mu = \frac{\partial E}{\partial n}$.  Then, the normalized chemical potential $\muh = \mu/\hbar\omega_z$
will be given by:
\be
\label{muhatehat}
\muh = \frac{1}{2}  \frac{\partial \hat{E}}{\partial n}.
\ee
To produce the curves in Fig.~\ref{fixedden}, then, we obtained $n$ and $\hat{E}$ as a function of the parameter $A$, which
can be combined to yield $\muh$ via Eq.~(\ref{muhatehat}) and hence $n$ as a function of $\muh$.  Note that the dimensionless 
density and coordinate comprising the vertical and horizontal axes of Fig.~\ref{fixedden} are identical to $n$ and $z$ of
this section (due to the abovementioned rescaling).  Thus, to implement the LDA, we obtain $n(z)$ from
\be
n(z) = n(\muh - \frac{1}{2}z^2),
\ee
with the function on the right being $n(\muh)$ as described above.  The central chemical potential in this formula is chosen to fix the
total particle number $N\simeq 250$ for each case.

\section{Concluding remarks}
 \label{sec:cr}
To conclude, although quasi 1D trapped Fermi gases are not expected to exhibit long-range
pairing order, short ranged pairing correlations will be induced by
the tunable attractive interactions and can be modeled by the simple variational
wavefunction Eq.~(\ref{eq:bcsform}).  Our theoretical approach, which does not rely on
the LDA (although it treats interaction effects approximately), 
can easily be implemented for experimentally realistic system parameters and, as shown here,
leads to specific predictions for how such pairing correlations impact
the momentum correlation function.   Since our approach agrees with the results of Bethe ansatz
plus LDA (at least in the weak coupling limit when the latter becomes more accurate), it provides
a simple description of trapped interacting fermionic atomic gases.

We gratefully acknowledge useful discussions with A. Chubukov,
R. Fernandes, F. Heidrich-Meisner, R. Hulet, A.M. Rey,  and I. Vekhter.  This
work was supported by the National Science Foundation Grant
No. DMR-1151717.  This work was supported in part by the National
Science Foundation under  Grant No. PHYS-1066293 and the hospitality
of the Aspen Center for Physics.  DES acknowledges support from the
German Academic Exchange Service (DAAD) and the hospitality of the 
Institute for Theoretical Condensed Matter physics at the Karlsruhe Institute of 
Technology.


\begin{thebibliography}{10}
\bibitem{BlochReview}
I. Bloch, J. Dalibard, and W. Zwerger, 
Rev. Mod. Phys. {\bf 80}, 885-964   (2008). 
\bibitem{Giorgini} S. Giorgini, L.P. Pitaevskii, and S. Stringari,
  Rev. Mod. Phys. \textbf{80}, 1215 (2008).
\bibitem{Guan}
X.-W. Guan, M.T. Batchelor, and C. Lee, Rev. Mod. Phys. 
{\bf 85}, 1633 (2013). 
\bibitem{FF} P. Fulde and R.A. Ferrell, Phys. Rev. \textbf{135}, A550
  (1964).  
\bibitem{LO} A.I. Larkin and Yu.N. Ovchinnikov, Zh. Eksp. Teor. Fiz
  {\bf 47}, 1136 (1964) [Sov. Phys. JETP {\bf 20}, 762 (1965)].
\bibitem{LiaoRittner}
Y. Liao, 
A.S.C. Rittner, T. Paprotta, W. Li, G.B. Partridge, R.G. Hulet, S.K. Baur, and E.J. Mueller, 
Nature (London) {\bf 467}, 567-569 (2010).
\bibitem{Orso07} G. Orso, Phys. Rev. Lett. \textbf{98}, 070402 (2007).
\bibitem{HuLiuDrummond} H. Hu, X.-J. Liu, and P.D. Drummond,
  Phys. Rev. Lett. {\bf 98}, 070403 (2007).
\bibitem{Radzihovsky} For a review see L. Radzihovsky and D.E. Sheehy,
  Rep. Prog. Phys. \textbf{73}, 076501 (2010).
\bibitem{RV} L. Radzihovsky and A. Vishwanath, Phys. Rev. Lett. {\bf 103}, 010404 (2009); L. Radzihovsky, Phys. Rev. A {\bf 84}, 023611 (2011).
\bibitem{Feiguin}
A. E. Feiguin and F. Heidrich-Meisner, Phys. Rev. B {\bf 76}, 220508 (2007).
\bibitem{LiuPRA2007} 
X.-J. Liu, H. Hu, and P. D. Drummond, Phys. Rev. A {\bf 76},
043605 (2007).
%

\bibitem{Batrouni} G. G. Batrouni,  M.H. Huntley, V.G. Rousseau, and R.T. Scalettar,
  Phys. Rev. Lett. \textbf {100}, 116405 (2008).
\bibitem{HM2010} F. Heidrich-Meisner, A.E. Feiguin, U. Schollw\"ock, and W. Zwerger, 
Phys. Rev. A {\bf 81}, 023629 (2010).
\bibitem{Sun} K. Sun,  J.S. Meyer, D.E. Sheehy, and S. Vishveshwara, 
Phys. Rev. A {\bf 83}, 033608 (2011).
\bibitem{SunBolech}
K. Sun and C.J. Bolech, Phys. Rev. A {\bf 85}, 051607 (2012). 
\bibitem{Altman04} E. Altman, E. Demler, and M.D. Lukin, 
Phys. Rev. A {\bf 70}, 013603 (2004).
\bibitem{Greiner05} M. Greiner, C.A. Regal, J.T. Stewart, D.S. Jin,
Phys. Rev. Lett. {\bf 94}, 110401 (2005).
\bibitem{Lu} H. Lu, L.O. Baksmaty, C.J. Bolech, and H. Pu, 
%
%
Phys. Rev. Lett. {\bf 108}, 225302 (2012).
\bibitem{Bolech12}
C.J. Bolech, et al, Phys. Rev. Lett. {\bf 109}, 110602 (2012).

\bibitem{DGpaper}
D.M. Gautreau, S. Kudla, and D.E. Sheehy, {\em in preparation}.
\bibitem{Olshanii98} M. Olshanii, Phys. Rev. Lett. \textbf{81}, 938
(1998).
\bibitem{Wang}  W.-M. Wang, Commun. Math. Phys. {\bf 277}, 459 (2008); W.-M. Wang,
%
 preprint http://arxiv.org/abs/0901.3970
\bibitem{Rey}
A.M. Rey, A.V. Gorshkov, C.V. Kraus, M.J. Martin, M. Bishof, M.D. Swallows, X. Zhang, C. Benko, J. Ye,
N.D. Lemke, and A.D. Ludlow, 
Annals of Physics. {\bf 340}, 311 (2014).
\bibitem{Gradshteynintegral}
The necessary integral is (for $s$ and $t$ integers)~\cite{Gradshteyn}:
%
\be
\int_{-\infty}^\infty dx\,  {\rm e}^{-2x^2}
H_{2s}(x) H_{2t}(x) = (-1)^{(s+t)}
2^{s+t-\frac{1}{2}} \Gamma\Big[s+t+\frac{1}{2}\Big].
\nonumber
\ee
\bibitem{Gradshteyn}
I.S. Gradshteyn and I.M. Ryzhik, Table of Integrals, Series, and Products.
\end{thebibliography}
\end{document}